\documentclass[runningheads]{llncs}

\usepackage{amsmath,amssymb}
\usepackage{xcolor}
\usepackage{url}
\usepackage{algorithmic}
\usepackage{algorithm}
\usepackage{paralist}
\usepackage{xspace}
\usepackage{graphicx}
\usepackage{tabularx}
\usepackage{subcaption}
\usepackage{multirow}
\usepackage{booktabs}
\usepackage{enumitem}
\usepackage[subtle]{savetrees}

\microtypesetup{
    final,
    tracking=true,
    kerning=true,
    spacing=true,
    expansion=true,
    protrusion=true,
    babel=true,
    factor=1000,
    stretch=10,
    shrink=10,
    step=1,
}

\makeatletter
\renewcommand\paragraph{%
  \@startsection{paragraph}{4}{\z@}%
                {0.5ex \@plus.5ex \@minus.2ex}%
                {-1em}%
                {\normalfont\normalsize\itshape}%
}
\makeatother

\def\BibTeX{{\rm B\kern-.05em{\sc i\kern-.025em b}\kern-.08em
    T\kern-.1667em\lower.7ex\hbox{E}\kern-.125emX}}

\newcommand{\fakeparagraph}[1]{\vspace{1mm}\noindent\textit{#1.}}

\newcommand{\fig}[1]{Fig.~\ref{fig:#1}}

\newcommand{\s}[1]{\S\ref{sec:#1}}

\newcommand{\eq}[1]{Eq.~\ref{eq:#1}}

\newcommand{\secref}[1]{\S\ref{sec:#1}}

\newcommand{\setN}{\mathcal{N}}
\newcommand{\setF}{\mathcal{F}}

\newcommand{\setC}{\mathcal{C}}
\newcommand{\setI}{\mathcal{I}}

\newcommand{\CSR}{\setC^{\mathrm{SR}}}
\newcommand{\CER}{\setC^{\mathrm{ER}}}
\newcommand{\CfSR}{\setC_f^{\mathrm{SR}}}
\newcommand{\CfER}{\setC_f^{\mathrm{ER}}}
\newcommand{\Fr}{\setF_{\mathrm{r}}}
\newcommand{\Fw}{\setF_{\mathrm{w}}}
\newcommand{\texec}{t^{\mathrm{exec}}}
\newcommand{\tretr}{t^{\mathrm{retr}}}
\newcommand{\Texec}{T_{\mathrm{exec}}}
\newcommand{\Tretr}{T_{\mathrm{retr}}}

\newenvironment{ceq}{%
  \begingroup
  \footnotesize
  \setlength{\abovedisplayskip}{3pt}%
  \setlength{\belowdisplayskip}{3pt}%
  \setlength{\abovedisplayshortskip}{2pt}%
  \setlength{\belowdisplayshortskip}{2pt}%
  \begin{equation}\begin{gathered}
}{%
  \end{gathered}\end{equation}%
  \endgroup
}

\newif\ifhidenotes
\hidenotestrue

\ifhidenotes
\newcommand{\noteam}[1]{}
\newcommand{\notedd}[1]{}
\else
\newcommand{\noteam}[1]{\footnote{\color{red}{Ale: #1}}}
\newcommand{\notedd}[1]{\footnote{\color{blue}{Dario: #1}}}
\fi

\newif\ifanonymous
\anonymousfalse

\begin{document}

\title{Replication-Aware Placement of Functions and Data in the Edge-Cloud Continuum}

\titlerunning{Replication-Aware Placement}

\ifanonymous

  \author{Anonymous Author(s)}
  \authorrunning{Anonymous Author(s)}

\else

  \author{
    Dario d'Abate\inst{1} \and
    Matteo Cenzato\inst{1} \and
    Matteo Briscini\inst{1} \and
    Arianna Dragoni\inst{1} \and
    Alessandro Margara\inst{1}
  }

  \authorrunning{D. d'Abate et al.}

  \institute{Politecnico di Milano, Milan, Italy \\
    \email{\{dario.dabate, arianna.dragoni, alessandro.margara\}@polimi.it}\\
    \email{\{matteo.cenzato, matteo.briscini\}@mail.polimi.it}}

\fi

\maketitle

\begin{abstract}
  Function-as-a-Service (FaaS) has emerged as the prominent programming
  model for the edge-cloud continuum. FaaS inherently decouples
  stateless functions from their persistent state.
  We study how to jointly schedule functions and place data to minimize client
  latency, considering data replication under heterogeneous consistency
  requirements.
  We introduce a Binary Linear Programming (BLP) model to compute optimal
  placements, establishing a rigorous theoretical baseline. Since the BLP
  scales cubically with the infrastructure nodes, we propose a
  topology-aware greedy heuristic that efficiently approximates the optimal
  solution.
  Our evaluation shows that the heuristic achieves near-optimal placement
  quality at a fraction of the computational cost, making it suitable for
  periodic system reconfigurations.
\end{abstract}

\section{Introduction}
\label{sec:intro}

Cloud computing has long been the de facto target for deploying applications,
offering managed environments with virtually unlimited resources. However, the
recent emergence of the edge-cloud continuum has fragmented the computing
infrastructure landscape~\cite{bittencourt:continuum:2025:compsci-review}.
Edge devices, deployed close to end users, enable lower latencies for
applications that are sensitive to response
times~\cite{shi:edge-comp:iot-journal:2016}. This is crucial for domains like
autonomous driving and industrial IoT, where milliseconds matter.
A dominant paradigm for the edge-cloud continuum is serverless
computing~\cite{nastic:serverless-edgetocloud:internet-comp:2017},
specifically Function-as-a-Service (FaaS), where managed runtimes handle the
scaling and placement of application functions across available nodes.
The FaaS model is inherently stateless: in cloud environments, when
applications require persistent state, this is delegated to external data
stores co-located in the same data center.
When functions execute at the edge, accessing a remote cloud-hosted store
incurs prohibitive latency, negating the benefits of edge
deployment~\cite{shi:edge-comp:iot-journal:2016}.
Removing this limitation gives rise to a challenging optimization problem. In
stateful serverless, we must decide not only where to execute each function
(\emph{function scheduling} problem), but also where to place the data items
that function accesses (\emph{data placement} problem).
Since multiple functions may access the same data items from different nodes,
the placement of each data item involves a trade-off: placing it close to one
consumer may penalize another. We refer to the combination of these two
decisions as the \emph{placement problem}.
Replication naturally mitigates this trade-off by placing data \emph{replicas}
on various nodes, enabling local access. However, replication introduces its
own challenges. Additional replicas reduce read latency by bringing data
closer to consumers, but updates must be propagated across all copies.
Furthermore, different applications impose different \emph{consistency
  requirements} on their data, ranging from strong guarantees that synchronize
all replicas before serving any read, to weaker guarantees that allow
replicas to diverge temporarily and converge over time. A placement
framework must therefore account for these consistency semantics, as they
fundamentally affect both the feasible placement space and the resulting
latency.
Existing approaches fail to solve this joint challenge. While some works
optimize scheduling without modeling data
placement~\cite{vahabi:energyEfficient,cicconetti:faasInIoT}, others couple
them but neglect variable consistency
requirements~\cite{nardelli:functionOffloading}. Furthermore, existing optimal
placement formulations using centralized solvers~\cite{baresi:neptune} suffer
from prohibitive computational overheads that preclude continuous or periodic
scaling. What is missing in the literature is a rigorous evaluation of an
optimal joint placement baseline under heterogeneous consistency models and an
efficient heuristic capable of approximating this optimum at runtime.

In this paper, we address the joint problem of function scheduling and data
placement in the edge-cloud continuum under heterogeneous consistency
requirements. Our contributions are as follows:
(1)~We formalize the problem and propose a Binary Linear Programming (BLP)
  model that jointly optimizes function scheduling and data replica placement
  under two consistency models: strong replication (SR), and eventual
  replication (ER).
(2)~We propose a topology-aware greedy heuristic that approximates the
  optimal solution efficiently, making it suitable for practical, periodic
  system reconfigurations.
(3)~We evaluate both approaches and show that the topology-aware heuristic
  achieves near-optimal placement quality compared to the BLP model, while
  reducing the computational cost required to find a solution.

The paper is organized as follows.
\s{related} reviews related work.
\s{system_model} formalizes the placement problem.
\s{centralized_mod} presents the BLP model.
\s{topology_aware_heuristic} describes the topology-aware greedy heuristic.
\s{eval} reports the evaluation results.
\s{conclusions} concludes the paper.

\section{Related Work}
\label{sec:related}

This section overviews existing approaches focusing on two dimensions: what
they model and how they make decisions.

\fakeparagraph{Function scheduling without data placement}
Several works optimize stateless execution in the edge-cloud continuum without
modeling data dependencies. Optimization approaches range from decentralized
heuristics using local routing logic (e.g., Cicconetti et
al.~\cite{cicconetti:faasInIoT}) to centralized solvers exploiting global
capabilities (e.g., Vahabi et al.~\cite{vahabi:energyEfficient} minimizing
energy consumption via ILP, or Baresi et al.~\cite{baresi:neptune} modeling
multi-access edge computing infrastructures via MILP). Rausch et
al.~\cite{rausch:containerScheduling}, integrate proximity scores into
orchestration frameworks but require manual data annotations.
None natively place data, which causes significant latency penalties in
stateful scenarios.

\fakeparagraph{Joint function scheduling and data placement}
Other works recognize that function scheduling and data placement are coupled,
but they either exclude replication or ignore different consistency semantics.
Nardelli et al.~\cite{nardelli:functionOffloading} jointly model function
offloading and data migration: the offloading part is decentralized, with each
node deciding based on local knowledge, while data migration assumes a
logically centralized migrator. They explicitly leave replication out of their
model.
Puliafito et al.~\cite{puliafito:stateAllocation} formulate a centralized MILP
that separates the problem into two sequential steps: first allocating
stateful microservices whose state is bound to the container, then dispatching
invocations to stateless serverless functions whose state remains in the
cloud. For the serverless part, data is not placed and replication is not
considered.
Smith et al.~\cite{smith:fado} build a platform that routes invocations to
clusters holding the required data buckets and supports replication across
clusters via asynchronous mirroring (MinIO), but the replication decisions are
manual and not automatically optimized.

\fakeparagraph{Consistency in stateful serverless}
A few platforms address consistency, but primarily at the protocol level
rather than a placement variable. Cloudburst~\cite{sreekanti:cloudburst}
leverages dynamic, back-pressure-driven replication and local mutable caches
over a key-value store.
Pfandzelter et al.~\cite{pfandzelter:enoki,pfandzelter:fred} integrate the
FReD replication middleware to offer multi-level consistency through optimistic
replication, but placement remains fully manual and replication targets all
designated nodes without cost optimization.
Other systems bypass consistency entirely via immutable data models
(Lambdata~\cite{tang:lambdata}) or treat state migration operationally via
locking (LoLa~\cite{wen:lola}).
In all cases, consistency constraints do not guide the systemic decision of
\emph{where} data and functions should be co-optimized.
\section{System Model and Problem Statement}
\label{sec:system_model}

\vspace{1mm}\noindent\textbf{Infrastructure.}
\label{sec:system_model:infrastructure}
We consider a set $\setN = \{n_1, \dots, n_N\}$ of heterogeneous nodes
spanning the edge-cloud continuum. Each node $n_i \in \setN$ is characterized
by its computational speed $\mathit{speed}_i$, memory capacity
$\mathit{mem}_i$, and storage capacity $\mathit{stor}_i$.

Edge-cloud deployments are usually arranged in layered tiers: conventionally
\emph{Cloud}, \emph{Fog}, and \emph{Edge}, where each layer may comprise
several logical levels~\cite{bittencourt:continuum:2025:compsci-review}.
Node density decreases from Edge to Cloud, and traffic typically aggregates
uplink toward the root; we therefore model the infrastructure as a
hierarchical tree where the Cloud is the root, a logical node with unlimited
resources, possibly representing a data-center cluster. The remaining nodes
form arbitrary subtrees rooted at the cloud, capturing deployments with any
number of intermediate tiers between Cloud and Edge.
Nodes are connected through the links of the tree. For every pair of nodes
$(n_i, n_j)$, we denote by $\mathit{lat}_{ij}$ the network latency between
them, computed as the sum of the latencies of the links along the unique path
connecting them in the tree, and by $\mathit{ban}_{ij}$ their effective
bandwidth, determined by the minimum bandwidth among the links on the same
path. We assume symmetric channels, so $\mathit{lat}_{ij} = \mathit{lat}_{ji}$
and $\mathit{ban}_{ij} = \mathit{ban}_{ji}$. Data retrieval from local storage
is considered negligible with respect to network latency.

\vspace{1mm}\noindent\textbf{Application Model.}
\label{sec:system_model:application}
Applications are composed of stateful serverless functions. Let $\setF =
  \{f_1, \dots, f_F\}$ denote the set of registered functions. Each function
$f$ is characterized by a reference execution time $\mathit{ref}_f$ and a
memory requirement $\mathit{mreq}_f$.
%
When a function is deployed on a node, to serve incoming requests, it incurs
this memory footprint $\mathit{mreq}_f$. In our model, this cost is paid only
once per node, regardless of the number of concurrent invocations.
This captures the shared, read-only memory components that are amortized
across multiple concurrent executions on the same host, such as the container
image layers, the runtime binary, and shared dynamic libraries.
End users (\emph{clients}) invoke functions from anywhere in the hierarchy;
the demand is captured by the invocation rate $\lambda_{f,i}$, which gives the
rate at which function $f$ is invoked from node $i$. We collect all pairs $(f,
  i)$ with $\lambda_{f,i} > 0$ in the set: $ \setI = \{(f, i) \in \setF \times
  \setN : \lambda_{f,i} > 0 \}$.
Scheduling is per pair $(f, i) \in \setI$, allowing invocations of the same
registered function from different sources to be routed independently; when
they converge on the same execution node, however, the function is deployed
there only once.
As data abstraction, we adopt the notion of \emph{collection}. A collection in
our model represents an opaque unit of state whose internal structure is
managed by the application or a higher-level middleware: a single collection
may wrap a composite data structure (e.g., documents, tables, nested records)
or a fine-grained record. What matters for the placement model is the
collection as the atomic unit of replication, not its internal layout.
Let $\setC = \{c_1, \dots, c_C\}$ denote the set of collections. Each
collection $c$ has a known size $\mathit{size}_c$ and a consistency strategy
$\mathit{sr}_c$, where $\mathit{sr}_c = 1$ denotes strong replication (SR) and
$\mathit{sr}_c = 0$ denotes eventual replication (ER).
Functions access the collections they require either locally, when co-located
on the same node, or remotely, by fetching data from another node at the cost
of additional network latency. The set of collections accessed by each
function, as well as the type of access (read or write), must be known
statically for a given problem instance; in practice, this information can be
derived from static annotations provided by the developer or automated code
analysis.
The access relation is modeled through the parameter $\mathit{access}_{c,f}$;
we write $\setC_f = \{c \in \setC : \mathit{access}_{c,f} = 1\}$ for the set
of collections accessed by function $f$. A function is classified as
\emph{read-only} ($\mathit{read}_f = 1$) if all its accesses are reads, and as
a \emph{writing function} ($\mathit{read}_f = 0$) if it performs at least a
write access.

\vspace{1mm}\noindent\textbf{Consistency Strategies.}
\label{sec:system_model:consistency}
Even within a single application, different collections may require different
consistency guarantees. Accordingly, each collection is associated with one of
two consistency strategies, which determine how it can be replicated and
accessed. We refer to collections managed under each strategy as \emph{SR
  collections} ($\mathit{sr}_c = 1$) and \emph{ER collections} ($\mathit{sr}_c =
  0$), respectively.

\fakeparagraph{Strong Replication (SR)}
Under SR, collections can be replicated across multiple nodes while preserving
a total order on write operations. We model this guarantee through a
single-leader protocol: one replica per collection is designated as the
\emph{leader} and serves all writes, while any replica can serve reads. The
leader propagates updates to followers asynchronously, outside the critical
path of function invocations. This is the most widely adopted design for
strong consistency in modern data stores.
We impose two constraints on functions that access SR collections. First, each
function accesses at most one SR collection (alongside any number of ER
collections). Second, a function that accesses an SR collection must execute
on a node that holds a replica of that collection; specifically, a writing
function must be co-located with the \emph{leader} replica, while a read-only
function can access any local replica.
These constraints are motivated by the prohibitive cost of cross-collection
coordination in geo-distributed environments. Traditional strong consistency
across independent collections requires global coordination protocols, which
impose synchronous barriers out of scale with edge-cloud latencies.
Restricting each function to a single SR collection eliminates this need, and
is in line with the design decisions of many distributed databases, such as
Cassandra. Moreover, leader co-location avoids remote synchronous round-trips.

\fakeparagraph{Eventual Replication (ER)}
Under ER, collections can be replicated without a designated leader. Any
replica can serve both reads and writes, with updates propagating
asynchronously and conflicts resolved through application-level policies
(e.g., last-writer-wins, CRDTs~\cite{shapiro:crdt:2011}). Because eventual
consistency does not enforce a total order on operations, there is no
transactional scope to bound: functions are free to access multiple ER
collections within a single invocation. Likewise, we relax the co-location
requirement: a function may execute without local copies of all its ER
collections, fetching missing ones from remote replicas at the cost of
additional retrieval latency.

\fakeparagraph{Fault tolerance}
At least one replica of each collection must persist in the cloud as a
recovery baseline. Beyond this, fault tolerance (node failures, leader
re-election) is outside the scope of this work.

\noindent\textbf{Placement Problem.}
\label{sec:system_model:problem}
Given the infrastructure, the set of registered functions with their
invocation statistics, and the set of collections with their consistency
strategies and access relations, the \emph{placement problem} consists of two
coupled decisions:
(1)~\emph{data placement}: on which nodes to place each collection and its
replicas, and
(2)~\emph{invocation scheduling}: on which node to execute each function
invocation originated from a source node.
These decisions are tightly connected: where collections are placed constrains
where functions can run, and where functions run determines the cost of
accessing remote collections (under ER).
The primary goal is to minimize the expected invocation latency, which
comprises the network delay between the client and the execution node, the
function execution time, and, under ER, the cost of fetching collections that
are not locally available. At the same time, replication must be kept under
control: placing replicas on every node would trivially eliminate remote
access latency, but at the cost of saturating the limited storage of edge
nodes. The placement must therefore balance data proximity against resource
consumption, subject to the finite memory and storage capacity of each node
and the co-location requirements imposed by the chosen consistency strategy.
\section{Centralized Model}
\label{sec:centralized_mod}

This section formalizes the placement problem (\secref{system_model}) as a
BLP. Our formulation handles both SR and ER collections: the $\mathit{sr}_c$
parameter activates the appropriate constraints and latency terms. For
convenience, we partition the collections accessed by function $f$ into $\CfSR
  = \{c \in \setC_f : \mathit{sr}_c = 1\}$ and $\CfER = \{c \in \setC_f :
  \mathit{sr}_c = 0\}$, and define the global sets $\CSR$ and $\CER$
analogously. As discussed in \secref{system_model:consistency}, we require
$|\CfSR| \leq 1$ for every $f$. Similarly, we partition functions into
read-only $\Fr = \{f \in \setF : \mathit{read}_f = 1\}$ and writing $\Fw = \{f
  \in \setF : \mathit{read}_f = 0\}$.

\vspace{1mm}\noindent\textbf{Decision Variables and Auxiliary Indicators.}
The formulation uses four families of primary binary decision variables,
encoding the placement, scheduling, leader election, and routing choices.
Specifically: $x_{c,j}=1$ iff a replica of $c$ is placed on $j$, $w_{f,i,j}=1$
iff $f$ from $i$ is scheduled on $j$, $l_{c,j}=1$ iff $j$ holds the leader of
$c$, and $r_{f,i,c,j,a}=1$ iff the invocation of $f$ from $i$, executing on
$j$, fetches $c$ from $a$.

\begin{ceq}
  x_{c,j} \in \{0,1\} \quad \forall\, c \in \setC,\; j \in \setN,\\
  w_{f,i,j} \in \{0,1\} \quad \forall\, (f,i) \in \setI,\; j \in \setN,\\
  l_{c,j} \in \{0,1\} \quad \forall\, c \in \CSR,\; j \in \setN,\\
  r_{f,i,c,j,a} \in \{0,1\} \quad \forall\, (f,i) \in \setI,\; c \in \CfER,\;
  j \in \setN,\; a \in \setN \setminus \{j\}.
\end{ceq}

Two binary auxiliary decision variables (hereafter \emph{auxiliary
  indicators}) track derived states and linearize logical dependencies:
$y_{f,j}=1$ iff at least one invocation of $f$ is scheduled on $j$, and
$z_{f,i,c,j}=1$ iff $f$ runs on $j$ and $c$ is not locally available.

\begin{ceq}
  y_{f,j} \in \{0,1\} \quad \forall\, f \in \setF,\; j \in \setN,\\
  z_{f,i,c,j} \in \{0,1\} \quad \forall\, (f,i) \in \setI,\; c \in \CfER,\; j
  \in \setN.
\end{ceq}

$y_{f,j}$ lets the memory capacity constraint (\eq{res}) charge
$\mathit{mreq}_f$ exactly once per node, while $z_{f,i,c,j}$ captures the
nonlinear product $w_{f,i,j}\cdot(1-x_{c,j})$, linearized via the McCormick
envelopes (\eq{er}).

\vspace{1mm}\noindent\textbf{Objective Function.}
%
The objective minimizes the average invocation latency, i.e., the mean latency
over all pairs $(f,i)\in\setI$, each weighted by its rate $\lambda_{f,i}$.
Since traffic is stationary, $\Lambda = \sum_{(f,i)\in\setI}\lambda_{f,i}$ is
constant, so minimizing $T/\Lambda$ reduces to minimizing the rate-weighted
total~$T$.

We formalize the placement objective as a lexicographic objective function
(\eq{lex}): first the model minimizes the rate-weighted total latency $T$,
then selects the placement with the lowest storage footprint $S$. Both are
normalized to $[0,1]$ using theoretical worst-case bounds $T^{\max}$ and
$S^{\max}$. The rate-weighted total latency $T$ decomposes into an execution
component $\Texec$ and a retrieval component $\Tretr$, both linear in the
decision variables $w$ and $r$:

\begin{ceq}
  \mathrm{lex\,min}\!\left(\tfrac{T}{T^{\max}},\, \tfrac{S}{S^{\max}}\right),
  \quad T = \Texec + \Tretr.
  \label{eq:lex}
\end{ceq}

\noindent
$\Texec$ aggregates per-invocation execution latency, weighted by the
corresponding rate $\lambda_{f,i}$ (\eq{texec}).
The formulation assumes a sequential execution model in which communication
and computation stages are not overlapped. Therefore, the per-invocation
latency $\texec_{f,i,j}$ is obtained by summing the round-trip network latency
between source node $i$ and execution node $j$ and the on-node computation
time:

\begin{ceq}
  \Texec = \textstyle\sum_{(f,i)\in\setI,\, j\in\setN}
  \lambda_{f,i}\,\texec_{f,i,j}\,w_{f,i,j},\\
  \texec_{f,i,j} = 2\,\mathit{lat}_{i,j} + \mathit{ref}_f/\mathit{speed}_j
  \quad \forall\,(f,i)\in\setI,\, \forall\,j\in\setN.
  \label{eq:texec}
\end{ceq}

Bandwidth is omitted: invocation payloads are assumed negligible compared to
the collection data transferred during retrieval.

\noindent
$\Tretr$ aggregates per-fetch retrieval latency, again weighted by
$\lambda_{f,i}$ (\eq{tretr}). The per-fetch cost is the round-trip latency
plus the transfer time of the collection over the available bandwidth, and
depends only on the link $(j,a)$ once $j$ is fixed:

\begin{ceq}
  \Tretr = \textstyle\sum_{(f,i)\in\setI,\, c\in\CfER,\, j\in\setN,\,
    a\in\setN\setminus\{j\}} \lambda_{f,i}\,\tretr_{c,j,a}\,r_{f,i,c,j,a},\\
  \tretr_{c,j,a} = 2\,\mathit{lat}_{j,a} + \mathit{size}_c/\mathit{ban}_{j,a}
  \quad \forall\,c\in\CER,\, \forall\,j\in\setN,\,
  \forall\,a\in\setN\setminus\{j\}.
  \label{eq:tretr}
\end{ceq}

This component is non-zero only for ER accesses: SR collections are guaranteed
locally available by the co-location constraints (\eq{sr}).

The worst-case rate-weighted latency $T^{\max}$ (\eq{tmax}) assumes, for each
$(f,i)$, the most expensive execution node and, for each ER collection, the
most expensive retrieval source.
The total storage cost $S$ (\eq{stor}) aggregates the size of all replicas
across all nodes, and its worst-case bound $S^{\max}$ corresponds to
replicating every collection on every node:

\begin{ceq}
  T^{\max} = \textstyle\sum_{(f,i)\in\setI}
  \lambda_{f,i}\,\max_{j\in\setN}\!\Big(\texec_{f,i,j} +
  \sum_{c\in\CfER}\max_{a\in\setN\setminus\{j\}}\tretr_{c,j,a}\Big),
  \label{eq:tmax}
\end{ceq}

\begin{ceq}
  S = \textstyle\sum_{c\in\setC,\, j\in\setN} \mathit{size}_c\,x_{c,j}, \quad
  S^{\max} = |\setN|\cdot\textstyle\sum_{c\in\setC} \mathit{size}_c.
  \label{eq:stor}
\end{ceq}

\vspace{1mm}\noindent\textbf{Constraints.}
Constraints fall into four groups.

\fakeparagraph{1. Structural constraints} Each invocation is scheduled on
exactly one node; the presence indicator $y_{f,j}$ activates iff $f$ has at
least one invocation on $j$; every collection has at least one replica:

\begin{ceq}
  \textstyle\sum_{j\in\setN} w_{f,i,j} = 1 \quad \forall\,(f,i)\in\setI,\\
  y_{f,j} \geq w_{f,i,j} \quad \forall\,(f,i)\in\setI,\, \forall\,j\in\setN,\\
  y_{f,j} \leq \textstyle\sum_{i:(f,i)\in\setI} w_{f,i,j} \quad
  \forall\,f\in\setF,\, \forall\,j\in\setN,\\
  \textstyle\sum_{j\in\setN} x_{c,j} \geq 1 \quad \forall\,c\in\setC.
  \label{eq:struct}
\end{ceq}

\fakeparagraph{2. SR-specific constraints} Each SR collection has exactly one
leader, co-located with a replica. Read-only invocations must execute where a
replica exists; writing invocations must execute on the leader:

\begin{ceq}
  \textstyle\sum_{j\in\setN} l_{c,j} = 1 \quad \forall\,c\in\CSR,\\
  l_{c,j} \leq x_{c,j} \quad \forall\,c\in\CSR,\, \forall\,j\in\setN,\\
  w_{f,i,j} \leq x_{c,j} \quad \forall\,(f,i)\in\setI \text{ with } f\in\Fr,\,
  \forall\,c\in\CfSR,\, \forall\,j\in\setN,\\
  w_{f,i,j} \leq l_{c,j} \quad \forall\,(f,i)\in\setI \text{ with } f\in\Fw,\,
  \forall\,c\in\CfSR,\, \forall\,j\in\setN.
  \label{eq:sr}
\end{ceq}

\fakeparagraph{3. ER-specific constraints} The miss indicator $z$ is
linearized via McCormick envelopes; on a miss, exactly one provider is
selected, and only nodes holding a replica can serve as providers:

\begin{ceq}
  z_{f,i,c,j} \leq w_{f,i,j},\quad z_{f,i,c,j} \leq 1 - x_{c,j},\\
  z_{f,i,c,j} \geq w_{f,i,j} + (1 - x_{c,j}) - 1,\\
  \textstyle\sum_{a\in\setN\setminus\{j\}} r_{f,i,c,j,a} = z_{f,i,c,j},\\
  r_{f,i,c,j,a} \leq x_{c,a},\\
  \forall\,(f,i)\in\setI,\, \forall\,c\in\CfER,\, \forall\,j\in\setN,\,
  \forall\,a\in\setN\setminus\{j\}.
  \label{eq:er}
\end{ceq}

\fakeparagraph{4. Resource constraints} Stored replicas must fit the per-node
storage limit; the memory footprint of deployed functions must fit the
per-node memory:

\begin{ceq}
  \textstyle\sum_{c\in\setC} \mathit{size}_c\,x_{c,j} \leq \mathit{stor}_j
  \quad \forall\,j\in\setN,\\
  \textstyle\sum_{f\in\setF} \mathit{mreq}_f\,y_{f,j} \leq \mathit{mem}_j
  \quad \forall\,j\in\setN.
  \label{eq:res}
\end{ceq}

\vspace{1mm}\noindent\textbf{Model Size.}
All constraints and both objective components are linear in the decision
variables, making the formulation a BLP with a lexicographic objective. The
routing variables $r_{f,i,c,j,a}$ dominate the model size with
$O(|\setI|\cdot|\CER|\cdot|\setN|^2)$ entries in the worst case. In practice
the number is significantly smaller: $z$ and $r$ are instantiated only when
$\mathit{access}_{c,f} = 1$ and $\mathit{sr}_c = 0$, and each function
typically accesses only a small subset of collections.
For further tractability, the number of replicas per collection can be bounded
by a constant $R_{\max}$ (e.g., $R_{\max} = 3$), reducing routing variables to
$O(|\setI|\cdot|\CER|\cdot|\setN|\cdot R_{\max})$:

\begin{ceq}
  \textstyle\sum_{j\in\setN} x_{c,j} \leq R_{\max} \quad \forall\,c\in\setC.
  \label{eq:rmax}
\end{ceq}

\section{Topology-Aware Heuristic}
\label{sec:topology_aware_heuristic}

While the BLP yields optimal solutions, its variable count grows as
$O(|\setI|\cdot|\CER|\cdot|\setN|^2)$, making it impractical for large
instances or periodic reconfiguration.
We propose a single-pass topology-aware greedy heuristic (hereafter \emph{TA})
that exploits the hierarchical tree infrastructure.
The algorithm involves four sequential phases, each producing a family of
decision variables. Feasibility is maintained by offloading excess storage and
memory demands up the tree to the root, which we model as a cloud with
virtually infinite resources.

\fakeparagraph{1. Collection placement}
We place at most $R_{\max}$ replicas per collection (with $R_{\max}$ the same
bound used in the BLP). For each collection $c$, we rank nodes by their
aggregate \emph{access pressure}, defined as the sum of invocation rates
$\lambda_{f,j}$ over all functions $f$ that read or write $c$. The top-ranked
nodes host the replicas; when a candidate lacks residual storage, the replica
is pushed up the tree until a node with sufficient capacity is found.
Collections with zero access pressure are placed directly at the cloud,
sparing the storage of lower nodes.

\fakeparagraph{2. Leader election}
Among the replicas of each SR collection, we elect as leader the one hosted by
the node with the highest aggregate \emph{write pressure} (the sum of
$\lambda_{f,j}$ over the writing functions accessing $c$). Co-locating the
leader with the heaviest writers eliminates a synchronous remote round-trip on
the most frequent updates. Ties are broken by node index for determinism.

\fakeparagraph{3. Function scheduling}
Each invocation $(f,i)\in\setI$ is assigned an execution node according to
$f$'s consistency requirements. If $f$ accesses an SR collection, the
co-location constraints (\eq{sr}) force execution on the nearest SR replica
(reads) or on the leader (writes); if that node lacks memory, the entire SR
cluster, the replica plus all invocations anchored to it, migrates upward
together to the first ancestor with enough capacity. If $f$ depends only on ER
collections, the algorithm targets the Lowest Common Ancestor (LCA) of the
nearest replicas of each required collection, walking up the tree if the LCA
is full. Functions with no data dependency are scheduled at the source node
$i$, again with upward fallback. The memory footprint of $f$ is charged at
most once per node, so repeated invocations of the same function on the same
node are free.

\fakeparagraph{4. Remote fetch routing}
For each ER access missing at the execution node, the provider is selected as
the nearest replica by hop count. 

\fakeparagraph{Complexity}
Each phase runs in polynomial time. The dominant cost is incurred by
scheduling and routing, which scan all nodes once per accessed collection for
every invocation. The overall complexity is $O(|\setI| \cdot |\setC| \cdot
  |\setN|)$, which is at most $O(|\setF| \cdot |\setC| \cdot |\setN|^2)$ since
$|\setI| \leq |\setF| \cdot |\setN|$.
\section{Evaluation}
\label{sec:eval}

We structure the evaluation around two research questions.

\smallskip\noindent \textbf{RQ1:} How do the problem dimensions (nodes,
collections, functions) and the consistency strategy affect the scalability of
the BLP in terms of solve time and memory footprint?  How does TA scale along
the same problem dimensions?

\smallskip\noindent \textbf{RQ2:} How closely does TA approximate the BLP
optimum in terms of total latency and total storage? How does it compare to
naive baselines?

RQ1 characterizes the scalability of the two approaches on the same workloads,
identifying where BLP becomes intractable and TA keeps producing solutions.
RQ2 evaluates, on instances where both complete, how closely TA approximates
the optimum, and quantifies the benefit of exploiting topology and data
locality against two naive baselines, defined later in this section.

\paragraph{Experimental Setup.}
\label{sec:eval-setup}

The BLP, TA, and the baselines are implemented in C++. The BLP uses the
Concert Technology of IBM ILOG CPLEX for model construction and solving.
All experiments were conducted on a server equipped with a 16-core AMD Ryzen 9
9950X processor (32 threads) and 64\,GB of DDR5 RAM, running Fedora Server 42.
A 600\,s timeout is enforced on the BLP solver, while virtual memory is capped
at 50\,GB.


We generate synthetic scenarios using a parametric workload generator. The
topology is a balanced $k$-ary tree whose root acts as the cloud and leaves as
edge nodes, with descending hierarchical capacities and ascending local
bandwidths.
%
%
%
Following \s{system_model:problem}, the cloud has unlimited resources as a
fallback, while edge nodes accommodate 40\% of the demand.
Network parameters are normally distributed: latency is $\mathcal{N}(10,
  2)$\,ms, bandwidth is $\mathcal{N}(100, 20)$\,Mbps, and node speed is
$\mathcal{N}(1.0, 0.2)$, with unconstrained local bandwidth ($10^9$\,Mbps).
Functions are invoked from $\mathcal{U}\{1, 3\}$ sources at $\mathcal{U}(1,
  10)$\,req/s, have a memory footprint of $\mathcal{N}(128, 32)$\,MB, an
execution time of $\mathcal{U}(1, 10)$\,ms, and a 0.8 read ratio. Finally,
collections have a size of $\mathcal{N}(50, 10)$\,MB, with each function
accessing $\mathcal{U}\{1, 3\}$ collections.
The dimensions $|\setN|$, $|\setF|$, $|\setC|$ and the consistency-related
parameters ($\mathit{srRatio}$, $\mathit{funcSrRatio}$, $R_{\max}$) are varied
systematically in subsequent subsections.

\paragraph{Scalability.}
\label{sec:eval-scalability}

To answer RQ1, we evaluate how BLP and TA scale as we independently vary the
number of nodes $|\setN|$, functions $|\setF|$, and collections $|\setC|$. For
each dimension we scale one parameter from~10 to $10^5$ while holding the
other two fixed at~10.
We further vary two orthogonal axes that affect the structure of the problem.
The first is the consistency mix of the workload: \emph{ER} (only
eventually-replicated collections), \emph{SR} (only strongly-replicated), and
\emph{MIX} (50/50). The second is the replication factor: \emph{no
  replication} ($R_{\max}=1$) and \emph{3 replicas} ($R_{\max}=3$). Each
configuration is solved with 10 different random seeds; we measure the median
and interquartile range (IQR) of total execution time (model construction plus
solving for BLP, single-pass execution otherwise) and maximum Resident Set
Size (RSS).

\begin{figure}[!t]
  \centering
  \includegraphics[width=\textwidth]{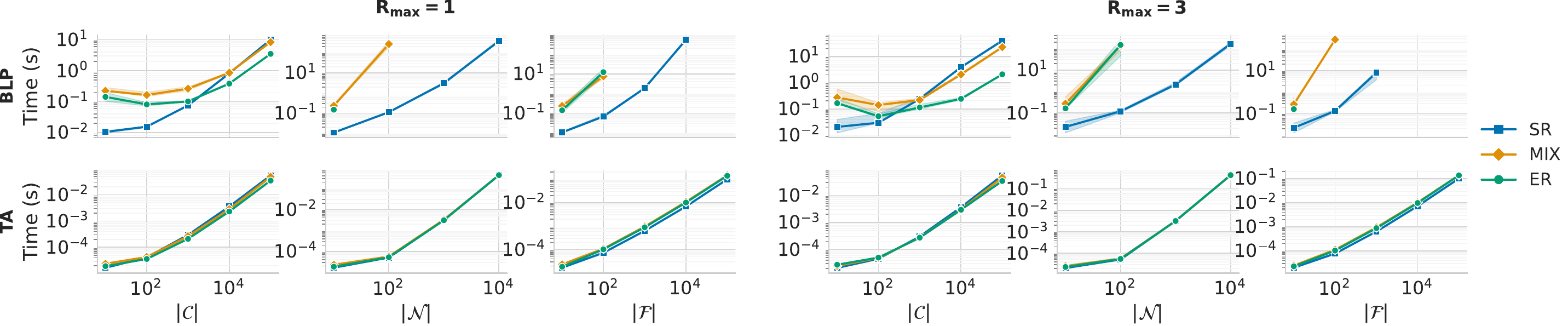}
  \caption{Scalability of BLP and TA.}
  \label{fig:scalability}
  \vspace{-4mm}
\end{figure}

\fig{scalability} reports the execution time of BLP (top) and TA (bottom)
scalability for both replication factors. Each panel plots the median over 10
seeds for ER, MIX, and SR, with vertical error bars indicating the
interquartile range (IQR); missing markers exceeded the 600\,s timeout or the
50\,GB memory cap.
\fig{scalability} shows that the BLP exhibits asymmetry across the three
scaling dimensions.
Collections are the most tractable dimension, as each function accesses at
most $\mathit{maxAccess}$ collections regardless of $|\setC|$, so the number
of active routing variables is bounded. Without replication, all three
consistency mixes complete $|\setC|=10^5$ within seconds and a few gigabytes
of RSS. Replication amplifies the cost, especially for MIX at $|\setC|=10^4$,
where the median execution time jumps to over $255$\,s, and at $|\setC|=10^5$
MIX exceeds the timeout.
When increasing the number of functions, ER and MIX time out already at
$|\setF|=10^3$ with no replication and at $|\setF|=10^2$ with three replicas,
while SR tolerates one to two orders of magnitude more ($|\setF|=10^4$ without
replication, $|\setF|=10^3$ with three replicas).
When scaling nodes, ER times out already at $|\setN|=10^2$ regardless of the
replication factor. Starting at $|\setN|=10^3$, both ER and MIX exhaust the
50\,GB memory cap.
%
This asymmetry reflects two distinct mechanisms: scaling $|\setN|$ inflates
the model directly (the ER routing variable $r_{f,i,c,j,a}$ grows cubically in
$|\setN|$), while scaling $|\setF|$ inflates the branch-and-bound search
space.
%
%
The execution time remains low across all dimensions: even the largest
instances we explored ($|\setF|=10^5$ or $|\setC|=10^5$) complete in well
under $150$\,ms with RSS below $150$\,MB, with only minor differences across
ER, MIX, and SR. Scaling the number of nodes is the most demanding regime: at
$|\setN|=10^4$ with three replicas TA completes all consistency mixes in about
$0.4$\,s using $2.5$\,GB of RSS, the cost being dominated by the size of the
pairwise hop-count matrix pre-computed for routing decisions. The contrast
with the BLP is most pronounced on the nodes dimension, where TA completes
$|\setN|=10^4$ for all mixes while the BLP exhausts the memory cap already at
$|\setN|=10^3$ for ER and MIX.
These results answer RQ1: BLP is suitable as an offline optimum reference at
small scale, is impractical for runtime use on realistic edge-cloud
infrastructures; TA, instead, bounds the search space and thus retains
acceptable cost across all the dimensions we explored.

\paragraph{TA Evaluation.}
\label{sec:eval-heuristic}

\begin{figure}[tpb]
  \centering
  \includegraphics[width=\textwidth]{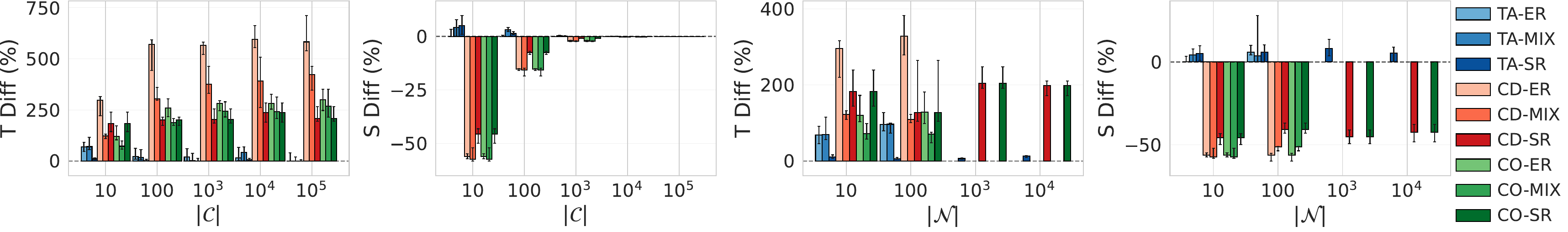}
  \caption{Solution quality comparison, 3 replicas ($R_{\max}=3$).}
  \label{fig:comparison_3rep}
  \vspace{-8mm}
\end{figure}

To answer RQ2, we evaluate the placement quality of TA against the BLP
optimum. We complement the comparison with two naive baselines that ignore
parts of the problem structure. For all approaches, we measure the
rate-weighted total latency $T$ and the total storage cost $S$.
A \emph{Cloud-only} (\emph{CO}) policy places a single replica of every
collection at the cloud, schedules every invocation there, and elects the
cloud as leader for all SR collections. It represents the canonical FaaS
deployment, with compute and data co-located within the data center.
A \emph{Cloud-data} (\emph{CD}) policy retains CO data placement but executes
ER-only invocations at the edge (walking upward in the tree to the first
ancestor with sufficient capacity if needed); SR invocations are forced to the
cloud by co-location, and every ER access from a non-cloud node triggers a
remote fetch from the cloud. CD captures the naive extension of FaaS to the
edge, where compute follows the user, but data stays centralized.
%
%
For each approach, we measure the percentage difference on $S$ and $T$
relative to the BLP optimum on the same instance.
%
We focus on two representative scenarios that bound the spectrum of problem
difficulty: scaling collections and scaling nodes, both with 3-replicas
factor. Each panel reports per-instance distributions over 10 seeds.
BLP-infeasible configurations are excluded.
\fig{comparison_3rep} reports the comparison when scaling collections (left)
and nodes (right): TA approaches the BLP optimum across the explored range,
while the relative cost of CO and CD depends on the consistency mix.
%
%
When scaling collections, the latency gap of TA stays small across consistency
mixes: TA-SR is within $10\%$ from the smallest scale and converges to zero
from $|\setC|=10^2$ onwards; TA-ER and TA-MIX medians start around $75\%$ at
$|\setC|=10$ and drop below $20\%$ from $|\setC|=10^2$, with a wider IQR
reflecting the variance introduced by ER routing decisions. The storage gap of
TA is within $10\%$ at small scale and collapses to zero as the workload
grows.
%
Both baselines use a single replica, so they consume less storage than BLP at
small scale, converging as $|\setC|$ grows. On SR, CD and CO coincide
(${\sim}200\%$ above BLP). On ER/MIX, CD lies above CO ($500$--$600\%$ vs.\
$200$--$300\%$): CD executes ER functions at the edge, saving the
client-to-compute hop, but each ER access triggers a cloud fetch, and the
cumulative cost of multiple fetches per invocation exceeds CO's single
centralized round-trip.
When scaling nodes the same ordering holds but with smaller magnitudes. The
number of comparable points is smaller, since the BLP only completes up to
$|\setN|=10^2$ for ER and MIX and up to $|\setN|=10^4$ for SR; further
configurations are excluded as discussed in \s{eval-scalability}. TA-SR
latency remains tightly bounded around $10\%$ above the BLP; TA-ER and TA-MIX
show a larger gap ($70$--$100\%$ at $|\setN|=10^2$) with wider IQR. TA storage
remains within $10\%$ across all mixes. CD and CO again coincide on SR
($\sim$$200\%$); on ER/MIX, CD stays in the $300$--$400\%$ range while CO is
  at $100$--$200\%$, confirming the pattern observed on collections.
These results answer RQ2: TA produces solutions whose latency and storage
remain within a small percentage of the BLP optimum on tractable instances. TA
also substantially outperforms both baselines. The CD-CO comparison further
shows that pushing computation to the edge without colocating data does not
exploit topology: only joint optimization of compute and data placement yields
the latency benefits of edge deployment.

\section{Conclusions}
\label{sec:conclusions}

We addressed joint function scheduling and data placement in the edge-cloud
continuum under heterogeneous consistency requirements, formulating a BLP and
proposing a fast topology-aware greedy heuristic.
Evaluation shows the BLP becomes intractable beyond a few hundred nodes, while
our heuristic scales gracefully, solving $10^5$-element instances in under
$150$\,ms. It produces solutions whose latency and storage are close to the
optimum, outperforming alternative baselines.
Our formulation targets a static placement for a stationary workload snapshot;
addressing runtime dynamics, along with distributed heuristics and validation
on real testbeds, is left to future work.

\bibliographystyle{splncs04}
\bibliography{biblio.bib}

\end{document}